\documentclass{egpubl}
\usepackage{EGVMV19}

\WsPaper           

\usepackage[T1]{fontenc}
\usepackage{dfadobe}

\usepackage{cite}  
\BibtexOrBiblatex
\electronicVersion
\PrintedOrElectronic
\ifpdf \usepackage[pdftex]{graphicx} \pdfcompresslevel=9
\else \usepackage[dvips]{graphicx} \fi

\usepackage{egweblnk}

\usepackage{times}
\usepackage{epsfig}
\usepackage{graphicx}
\usepackage{amsmath}
\usepackage{amssymb}

\usepackage{algorithm}
\usepackage{algpseudocode}
\usepackage{wrapfig}
\usepackage{subcaption}

\usepackage{caption}
\usepackage{array}
\usepackage{epstopdf}
\usepackage{cuted}
\begingroup
  \catcode`\_=\active
  \gdef_#1{\ensuremath{\sb{\mathrm{#1}}}}
\endgroup
\mathcode`\_=\string"8000
\catcode`\_=12

\usepackage{xcolor}
\usepackage{textcomp}
\definecolor{Gray}{rgb}{0.5,0.5,0.5}
\definecolor{darkblue}{rgb}{0,0,0.7}
\definecolor{orange}{rgb}{1,.5,0} 
\definecolor{red}{rgb}{1,0,0} 

\newcommand{\Matrix}[1]     {{\ensuremath{\mathbf{\uppercase{#1}}}}} 
\newcommand{\Vector}[1]     {{\ensuremath{\mathbf{\lowercase{#1}}}}} 

\newcommand{\mini}[1]   {\underset{{#1}}{\operatorname{min}} \: \: } 
\newcommand{\minimize}[1]   {\underset{{#1}}{\operatorname{argmin}} \: \: } 

\newcommand{\signal} {\Vector{x}}
\newcommand{\filter} {\Vector{d}}
\newcommand{\code}   {\Vector{z}}
\newcommand{\Filter} {\Matrix{D}}
\newcommand{\mask}   {\Matrix{M}}
\newcommand{\Code}   {\Matrix{z}}
\newcommand{\surC}   {\Matrix{C}}
\newcommand{\surB}   {\Matrix{B}}
\newcommand{\matA}   {\Matrix{A}}
\newcommand{\ite}	 {t}
\newcommand{\subcode}{\tilde{\code}}

\title{Stochastic Convolutional Sparse Coding}

\author[J. Xiong \& P. Richtarik \& W. Heidrich]
{\parbox{\textwidth}{\centering J. Xiong\orcid{0000-0002-1507-6070},
		P. Richtarik\orcid{0000-0003-4380-5848}
		and W. Heidrich\orcid{0000-0002-4227-8508}
       }
        \\
{\parbox{\textwidth}{\centering King Abdullah University of Science and Technology (KAUST), Saudi Arabia
       }
}
}

\begin{document}

\maketitle
\begin{abstract}
  State-of-the-art methods for Convolutional Sparse Coding usually
  employ Fourier-domain solvers in order to speed up the convolution
  operators. However, this approach is not without shortcomings. For
  example, Fourier-domain representations implicitly assume circular
  boundary conditions and make it hard to fully exploit the sparsity
  of the problem as well as the small spatial support of the filters.
  
  In this work, we propose a novel stochastic spatial-domain solver,
  in which a randomized subsampling strategy is introduced during the
  learning sparse codes. Afterwards, we extend the proposed strategy
  in conjunction with online learning, scaling the CSC model up to
  very large sample sizes. In both cases, we show experimentally that
  the proposed subsampling strategy, with a reasonable selection of
  the subsampling rate, outperforms the state-of-the-art
  frequency-domain solvers in terms of execution time without losing
  the learning quality. Finally, we evaluate the effectiveness of the
  over-complete dictionary learned from large-scale datasets, which demonstrates an
  improved sparse representation of the natural images on account of
  more abundant learned image features.


\end{abstract}  

\section{Introduction}
Convolutional Sparse Coding (CSC) is a method for learning {\em
  generative} models in the form of translationally invariant
dictionaries for a large variety of different training signals.  These
generative models have been shown effective for solving problems in
neural and brain information
processing~\cite{jas2017learning,peter2017sparse}, as well as in a
variety of image processing tasks, for instance, image
inpainting~\cite{heide2015fast},
super-resolution~\cite{gu2015convolutional}, high dynamic range
imaging~\cite{serrano2016convolutional}, and high-dimensional signal
reconstructions~\cite{choudhury2017consensus,bibi2017high}. CSC
differs from conventional sparse coding by formulating the signals as
the sum of a set of convolutions on dictionary filters and sparse
codes instead of patch-wise linear combinations of
filters. In traditional sparse dictionary learning, the patch
structure significantly degrades the expressiveness of the
dictionaries by introducing a strong dependency on the position of a
feature, which the convolutional nature of CSC avoids.

This convolutional approach is also at the heart of many deep
learning-based methods in the form of
CNNs~\cite{lecun1998gradient,kavukcuoglu2010learning,krizhevsky2012imagenet},
which have in recent years been extraordinarily successful for a broad
range of high-level image understanding applications. However, while
CNNs generally are used in a {\em supervised} setting and produce {\em
  discriminative}, task-specific models, CSC is {\em unsupervised} and
produces {\em generative} models that can easily be transferred between tasks.

To solve the optimization problems inherent to CSC, Zeiler et
al.~\cite{zeiler2010deconvolutional} iteratively solve two subproblems
(updating sparse codes and updating filters) using gradient decent in
the form of convolutional operations in the spatial domain, which is
computationally expensive. Recent algorithms tackle the problem by
exploiting Parseval's theorem to express the spatial convolution by
multiplication in the frequency domain and using proximal solver such
as Alternating Direction Method of Multipliers
(ADMM)~\cite{boyd2011distributed} to separate the linear least squares
parts from the non-smooth terms in the optimization
problem. These approaches show tremendous improvements over prior
spatial-domain solvers with respect to running
time~\cite{bristow2013fast,heide2015fast,wohlberg2016efficient,choudhury2017consensus}. Most
of the prior work learns the dictionary filters in a batch mode, which
indicates that all training signals are involved in every training
iteration, and this restricts it from applying to large datasets or
streaming data.

In contrast to batch mode learning, online
learning~\cite{shalev2012online} is a well established strategy which
processes a single image or a small portion (mini-batch) of the whole
data at each training step, and incrementally updates model
variables. Herein, the required memory and computing sources are only
dependent on the sample size in every observation, independent of the
training data size. It alleviates the scalability issue that arises in
batch approaches, and the convergence of the algorithm was firstly
analyzed using stochastic approximation
tools~\cite{bottou1998online}. Bottou et
al.~\cite{bousquet2008tradeoffs} further showed better generalization
performance of the stochastic algorithms than standard gradient
descent on large scale learning systems. Later on, online learning
strategies were synergetic with sparse coding, which was then scaled
up for learning dictionary from millions of training
samples~\cite{mairal2009online,mairal2010online}, and for large-scale
matrix factorization with an additionally introduced subsampling
strategy~\cite{mensch2016dictionary}. More recently, Liu et
al.~\cite{liu-2018-first} and Wang et al.~\cite{wang2018scalable}
separately proposed similar online learning frameworks for the CSC
model, alleviating the memory issues arise in batch-based CSC model on
large datasets.

{\bfseries Contributions.} We mainly make three contributions in this
work. First, we introduce a randomization strategy for the CSC
model and solve the entire problems in the spatial domain. We demonstrate
that the proposed stochastic spatial-domain solver, with a reasonably
selected subsampling rate, outperforms the state-of-the-art
frequency-domain solvers with regard to computing efficiency. Secondly, we
formulate an online-learning version of the proposed algorithm, and
show dramatic runtime improvement over current online CSC methods,
while producing comparable outcomes. Finally, we demonstrate the
capability to learn the meaningful over-complete dictionary from thousands of
images, and analyze the effectiveness of the learned over-complete
dictionary for a number of reconstruction tasks.


\section{Convolutional Sparse Coding (CSC)} \label{CSCmodel}
The dictionary learning problem for CSC problem has the form
\begin{equation}\label{eq:CSCmodel}
\begin{split}
    \mini{\filter,\code} & \frac{1}{2}\|\signal - \sum_{k=1}^{K} \filter_k * \code_k \|_2^2 + \lambda \sum_{k=1}^{K}\| \code_k \|_1 \\
    \text{subject to} & ~ \|\filter_k\|^2_2 \leq 1 ~~ \forall k \in \{1,\dots,K\},
\end{split}
\end{equation}
where $\signal \in \mathbb{R}^D$ is a $D$-dimensional signal or a
vectorized image,
$\filter_k \in \mathbb{R}^M$ is the $k$-th dictionary, $\code_k\in
\mathbb{R}^D$ is the sparse code associated with that dictionary,
$\lambda>0$ is a sparsity inducing penalty parameter, $K$ is the
number of dictionary filters, and $*$ is the convolution operator. The
above model will be applied to all the training images
$\signal\in\mathbb{X}$.

Most recent CSC algorithms exploit Parseval's theorem and introduce
two slack variables to separate the non-smooth $L_1$ penalty term and
the $L_2$ constraints, making it feasible to efficiently compute the
latter in the frequency domain. Furthermore, the whole
Problem~\eqref{eq:CSCmodel} can be split into alternating subproblems
for updating $\code$ and $\filter$, which are jointly solved by
coordinate
descent~\cite{bristow2013fast,heide2015fast,wohlberg2016efficient}. This
approach suffers from several issues:

\begin{itemize}
  \item While CSC overcomes the independence assumption held in
    patch-based learning algorithms, far more variables ($K$ times
    more) are introduced to represent a single image to compensate for
    this. This creates more severe memory and computational burdens.

  \item We observe through experiments that the vast majority of the
    entries of the reconstructed sparse codes do not provide useful
    information about the represented image. For $K=100$, 99.5\%
    entries are not informative. This indicates that the subproblem
    for updating $\code$ solves a highly sparse LASSO
    problem. Transforming the problem into frequency domain imposes
    restrictions on exploiting this sparsity.

  \item While prior work shows its efficiency in solving the CSC
    problem in the frequency domain, this is only applicable for
    updating $\code$, and does not hold for updating $\filter$. The
    dictionary filters usually have much smaller spatial support than
    the dimension size of the sparse codes ($M \ll D$). However, in
    order to tackle the problem in the frequency domain, it is
    necessary to process the $\filter$-subproblem over the full
    support of the sparse codes, and then project the results onto the
    much smaller spatial support of the filters.
\end{itemize}


\section{Stochastic Convolutional Sparse Coding}
\subsection{The SCSC Model}
We first define a stacked vector $\code = [\code_1, \dots, \code_K]
\in \mathbb{R}^{DK}$ for the codes, as well as a stacked matrix
$\Filter = [\Filter_1, \dots, \Filter_K] \in \mathbb{R}^{D \times DK}$
for the filter convolutions. The convolution operators are expressed
as a matrix multiplication so that $ \Filter \code = \sum_{k=1}^{K}
\filter_k * \code_k$.  Therefore, each part of $\Filter$ is a Toeplitz
matrix.

Based on the strong sparsity of the codes $\code$, we propose to
implement the CSC learning problem~\eqref{eq:CSCmodel} iteratively,
where at the $\ite$-th iteration we only consider a random subset of
the codes, denoted as $\subcode^t\in \mathbb{R}^{pDK}$.  We subsample
the sparse codes $\code$ following a Bernoulli distribution with
probability $p$. This subsampling process can be expressed as a matrix
operation
\begin{equation}
    \subcode^t = \mask^{\ite} \code^\ite,
\end{equation}
where $\mask^{\ite}$ is a $\mathbb{R}^{pDk \times DK}$ binary matrix,
where exactly one entry per row has a value of $1$, and the other
entries are $0$. For each iteration $\ite$, a different matrix
$\mask^{\ite}$ is generated randomly. This matrix projects the
codes $\code^\ite$ to a random subspace, retaining the sampled codes
and filtering out the others. In the case of $p=1$, the proposed model
is identical to the classical CSC model, and $\mask$ is simply an
identity matrix. When $p<1$, the algorithm only solves a subset of the
codes at chosen positions in each iteration, and accordingly, the
update of dictionary $\filter$ is based on the selected portion of the
codes $\subcode$. Similar to solving the classical CSC problem, we can
apply a coordinate descent algorithm, alternating on subproblems of
$\subcode$ and $\filter$, to tackle the bi-convex optimization
problem. Specifically, the modified minimization problem for
$\subcode$ can be formulated as:
\begin{equation} \label{eq:updatingCode}
    \subcode^{\ite} = \minimize{\subcode} \frac{1}{2}\| \signal - \Filter^{\ite-1} (\mask^{\ite})^\top \subcode \|_2^2 + \lambda \| \subcode \|_1,
\end{equation}
where $\Filter^{\ite-1}$ is composed of the dictionary learned in the
$(\ite-1)$-th iteration.
 Due to the introduced subsampling matrix, the convolution
operator cannot be implemented in the Fourier domain. However, owing
to the subsampling strategy, the number of variables that need to be
computed for this subproblem is $pDK$ instead of $DK$, which leads to
a reduction of the spatial-domain computation time by a factor of
$\frac{1}{p}$.

After computing the subsampled codes $\subcode^\ite$, we can then project them
onto the original spatial support by
\begin{equation} \label{eq:upsampleCode}
    \code^\ite = (\mask^{\ite})^\top \subcode^\ite.
\end{equation}
Afterwards, the dictionary can be updated by solving the optimization problem
\begin{equation} \label{eq:updatingFilter}
\begin{split}
   & \filter^{\ite} = \minimize{\filter} \frac{1}{2}\|\signal - \Code^{\ite} \filter \|_2^2 \\
   & \text{subject to}  ~ \|\filter_k\|^2_2 \leq 1 ~ \forall k \in \{1,\dots,K\},
\end{split}
\end{equation}
where $\Code^{\ite}= [\Code_1^{\ite}, \dots, \Code_K^{\ite}]
\in \mathbb{R}^{D \times MK}$ is a concatenation of Toeplitz matrices,
and $\Code_k^{\ite}$ is constructed from the associated $\code_k^{\ite}$, $\filter=
[\filter_1,\dots,\filter_K] \in \mathbb{R}^{MK}$ such that $ \Code
\filter = \sum_{k=1}^{K} \filter_k * \code_k$. In typical CSC
settings, $M \ll D$, hence there is no need to perform subsampling on
the dictionary. Notice that the time complexity of solving the $\filter$-update 
step in spatial domain is dependent of $M$, in contrast with that of frequency-domain 
solvers, which is $D$ dependent. This addresses the third issue in Section~\ref{CSCmodel}.

Our proposed randomization approach utilizes ideas similar to the stochastic Frank-Wolfe
algorithm~\cite{reddi2016stochastic,pmlr-v80-kerdreux18a}.  The general idea is to solve the optimization problem on a subset of the variables at each iteration, which are
randomly extracted based on a certain probability distribution. For the proposed algorithm, each iteration extracts
($pDK+MK$) variables, where $pDK$ variables are randomly picked from a
total $DK$ variables, and the rest remain unchanged. Owing to the fact
that CSC model is over-parameterized and the codes are highly 
sparse, the original signals can still be represented by a portion of 
the codes under a reasonable subsampling
rate. Therefore, the convergence of the proposed algorithm will not be
significantly affected by the subsampling manipulation, unlike the
general case of the stochastic Frank-Wolfe algorithm, which usually
requires more iterations to reach convergence. This insight is
experimentally verified in Section~\ref{sec:result}.

In the following we introduce two different outer loop structures to
utilize the proposed subsampling strategy. First, we introduce a {\em
  batch mode} method (stochastic batch CSC) that learns from all
images simultaneously, and second we introduce an {\em online}
variant (stochastic online CSC).

\subsection{Stochastic Batch CSC (SBCSC)}
We first introduce a batch-mode version of the proposed method as
shown in Algorithm~\ref{algo:SBCSC}, where $N$ is the number of total
input images, $(\subcode^i)^{\ite}$ is the sampled codes associated with $i$-th
image at $\ite$-th iteration and $(\code^i)^{\ite}$ is the corresponding codes in the original spatial support, $p$ is the uniform probability for one code been
selected. We choose $p=\{1, 0.5, 0.2, 0.1, 0.05\}$ for testing in this
work, where $p=1$ indicates no subsampling, and $p=0.05$ indicates a
subsampling rate of $5\%$.

\begin{algorithm}[H]
\caption{SBCSC} \label{algo:SBCSC}
\begin{algorithmic}[1]
\State $\text{Initialize}  ~ \ite=0, ~\filter^t, ~p$
\While {not converge}
    \State $\ite \gets \ite+1$
    \State $ \text{Randomly sample }\code^{\ite} \text{ with rate } p $
    \For{i=1 to N}
        \State $ \text{Compute } (\subcode^i)^{\ite} \text{ by solving problem~(\ref{eq:updatingCode}})$
        \State $ \text{Compute } (\code^i)^{\ite}   \text{ by Eq.~\ref{eq:upsampleCode}}$
    \EndFor
    \State $\text{Compute } \filter^{\ite} \text{ by solving problem~(\ref{eq:updatingFilter}})$
\EndWhile
\end{algorithmic}
\end{algorithm}

Problem~\eqref{eq:updatingCode} is the standard LASSO, which can be
solved by plenty of optimization frameworks. We found that solving it
with ADMM delivers a good balance between computation time and
convergence within a moderate number of iterations. Specifically, the
data fitting term and the $L_1$ penalty term are split, forming two
separate substeps. The first substep is a quadratic programming (QP)
problem, and we can either cache the matrix factorization by Cholesky
decomposition (when $N$ is relatively large), or solve it iteratively
by Conjugate Gradient (when $N$ is relatively small). The second
substep can be solved by a point-wise shrinkage operation.
Problem~\eqref{eq:updatingFilter} is a quadratic constrained
quadratic programming (QCQP) problem, and it can be efficiently solved
by projected block coordinate decent. Empirically, a single iteration
is enough with $\filter$ computed in previous iteration as a warm
start. We set the hyper-parameters $\lambda=1$, the ADMM iteration
fixed to 10, the augmented Lagrangian penalty $\rho$ to $10 \lambda$,
and the over-relaxation strategy within ADMM is applied with $\alpha =
1.8$. For a detailed description of the above two solvers, please
refer to the supplement.

Every outer loop of SBCSC involves all of the training images, which
makes it computationally expensive to process them all simultaneously.
Furthermore, memory consumption quickly becomes an issue with
increasing numbers of images. Thus, the batch-based learning algorithm
lacks the ability to scale up to very large datasets or to handle
dynamically changing training data.

\subsection{Stochastic Online CSC (SOCSC)}

\begin{algorithm}[H]
\caption{SOCSC} \label{algo:SOCSC}
\begin{algorithmic}[1]
\State $\text{Initialize} ~ t=0, ~\filter^t, ~p, ~\surC^t = 0, ~\surB^t = 0$
\While {not converge}
    \State $t \gets t+1$
    \State $ \text{draw } \signal^t \text{ from training images} $
    \State $ \text{Randomly sample }\code^t \text{ with rate } p $
    \State $ \text{Compute } \subcode^t \text{ by solving problem~(\ref{eq:updatingCodeOnline}) using } \filter^{t-1}$
    \State Compute $\code^t$ by Eq.~\ref{eq:upsampleCode}
    \State Compute $\surC^t$ and $\surB^t$ by Eq.~\ref{eq:updateSur}
    \State Compute $\filter^t$ by solving problem~(\ref{eq:updatingFilterOnline})
\EndWhile
\end{algorithmic}
\end{algorithm}

In order to address this scalability issue, we can tackle the
Stochastic CSC problem in an online fashion. In the online learning
setting as shown in Algorithm~\ref{algo:SOCSC}, each iteration only
draws one or a subset (mini-batch) of the total training images, hence
the complexity per loop is independent of the training sample
size. Then, given the sampled image $\signal^t$ at $t$-th iteration,
we can compute the corresponding subsampled sparse codes $\subcode^t$ by
\begin{equation} \label{eq:updatingCodeOnline}
    \subcode^t = \minimize{\subcode} \frac{1}{2}\|\signal^t - \Filter^{t-1} (\mask^t)^\top \subcode \|_2^2 + \lambda \|\subcode\|_1.
\end{equation}
The only difference between Eq. ~\ref{eq:updatingCode} is that $\signal^t$ only contains a portion of the total images. After projecting $\subcode^t$ onto the original spatial support and obtaining the sparse codes $\code^t$, the dictionary is updated by:
\begin{equation}
\begin{split}
    &\filter^t = \minimize{\filter} \frac{1}{2t}\sum_{i=1}^{t} \|\signal^i - \Code^i \filter \|_2^2 \\
    &\text{subject to} \quad \|\filter_k\|^2_2 \leq 1 ~ \forall k \in \{1,\dots,K\}.
\end{split}
\end{equation}
Note that updating the dictionary in this fashion involves all of the
past training images and sparse codes. Using techniques developed for
regular (non-convolutional) dictionary
learning~\cite{mairal2009online,mairal2010online}, we can get rid of
explicitly storing this data by introducing two surrogate matrices
$\surC \in \mathbb{R}^{KM \times KM}$ and $\surB \in \mathbb{R}^{KM
  \times 1}$. These carry all of the required information for updating
$\filter$, and can be iteratively updated by:
\begin{equation} \label{eq:updateSur}
\begin{split}
    \surC^t  = \frac{t-1}{t} \surC^{t-1} + \frac{1}{t}(\Code^t)^\top \Code^t \\
    \surB^t  = \frac{t-1}{t} \surB^{t-1} + \frac{1}{t}(\Code^t)^\top \signal^t
\end{split}
\end{equation}
With these surrogate matrices, the updated dictionary can be obtained
by solving
\begin{equation} \label{eq:updatingFilterOnline}
\begin{split}
    & \filter^t = \minimize{\filter} \frac{1}{2} \filter^\top \surC^t\filter - \filter^\top \surB^t \\& \text{subject to} \quad \|\filter_k\|_2^2 \leq 1 ~ \forall k \in \{1,\dots,K\}.
\end{split}
\end{equation}
Problem~\eqref{eq:updatingCodeOnline} and
problem~\eqref{eq:updatingFilterOnline} are solved in the same way as
that for SBCSC.

\subsection{Complexity Analysis}\label{complexity}
Recall that $D$ is the number of pixels for a single image, $K$ is the
number of filters, and $M$ is the size of the filter
support. Commonly, we can assume $K \approx M$.  State-of-the-art
frequency-domain solvers then have the time complexity
$\mathcal{O}(K^2D + KDlog(D))$ for a single data pass.

The time complexity of updating $\code$ in our SBCSC algorithm using
Conjugate Gradient is $\mathcal{O}(pKMD \sqrt{\tau})$, where $pKMD$ is
the number of non-zero elements in $(\Filter \mask^\top)$ and $\tau$
is the condition number of $(\matA^\top \matA + \rho I)$ where $\matA
= \Filter \mask^\top$. With a reasonable selection of the subsampling
rate, this time complexity is comparable to that of frequency-domain
solvers.

Updating the filters $\filter$ takes $\mathcal{O}(K^2M^2)$ time. This
is comparable to $\mathcal{O}(K^2D)$ in the common CSC setting ($M \ll
D$). However, multiple ADMM iterations are required in the frequency
domain to compute $\filter$ while only a single pass is required by
the proposed method, which greatly reduces the computation
time. Overall, the proposed method has the time complexity of
$\mathcal{O}(pKMD \sqrt{\tau} + K^2M^2)$.

The time complexity of SOCSC for one data pass is similar to that of SBCSC, apart from
two additional steps to update the surrogate matrices. Updating
$\surC$ and $\surB$ involves computing $\Code^\top\Code$ and
$\Code^\top x$. Although $\Code$ has dimensions of $D \times KM$, it
is a highly sparse matrix with only $\mathcal{O}(D)$ non-zero
elements. Therefore, the total performance is not affected
significantly.


\section{Experiments and Results} \label{sec:result}

 \subsection{Experimental Design}
We first validate the proposed algorithms on the fruit and city
datasets~\cite{zeiler2010deconvolutional}, that each consists of 10
training images of size $100 \times 100$. The online-mode algorithms
are then adapted to one thousand $100 \times 100$ image patches
randomly picked up from ImageNet~\cite{deng2009imagenet}. Note that batch-based CSC commonly can only handle less than 100 images
simultaneously due to the memory limitation. The dictionary size is set to $100$ filters of size $11 \times 11$ pixels in all experiments except for over-complete dictionary. All training and evaluation processes in this manuscript are performed on contrast normalized
images~\cite{zeiler2010deconvolutional,heide2015fast}.

\begin{figure}[h]
\centering
\begin{subfigure}{0.45\textwidth}
  \includegraphics[width=1\linewidth]{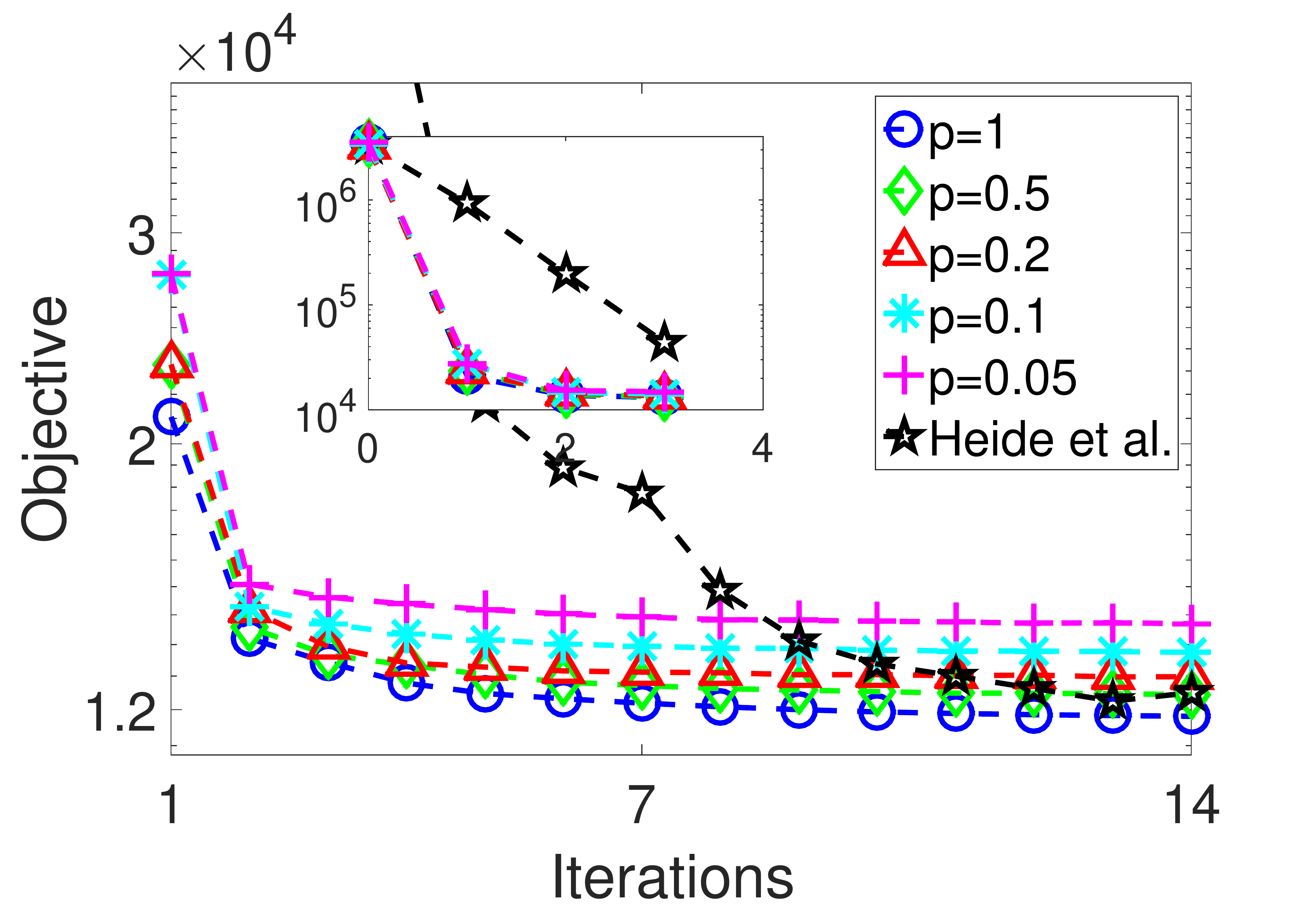}
\end{subfigure}

\begin{subfigure}{0.45\textwidth}
  \includegraphics[width=1\linewidth]{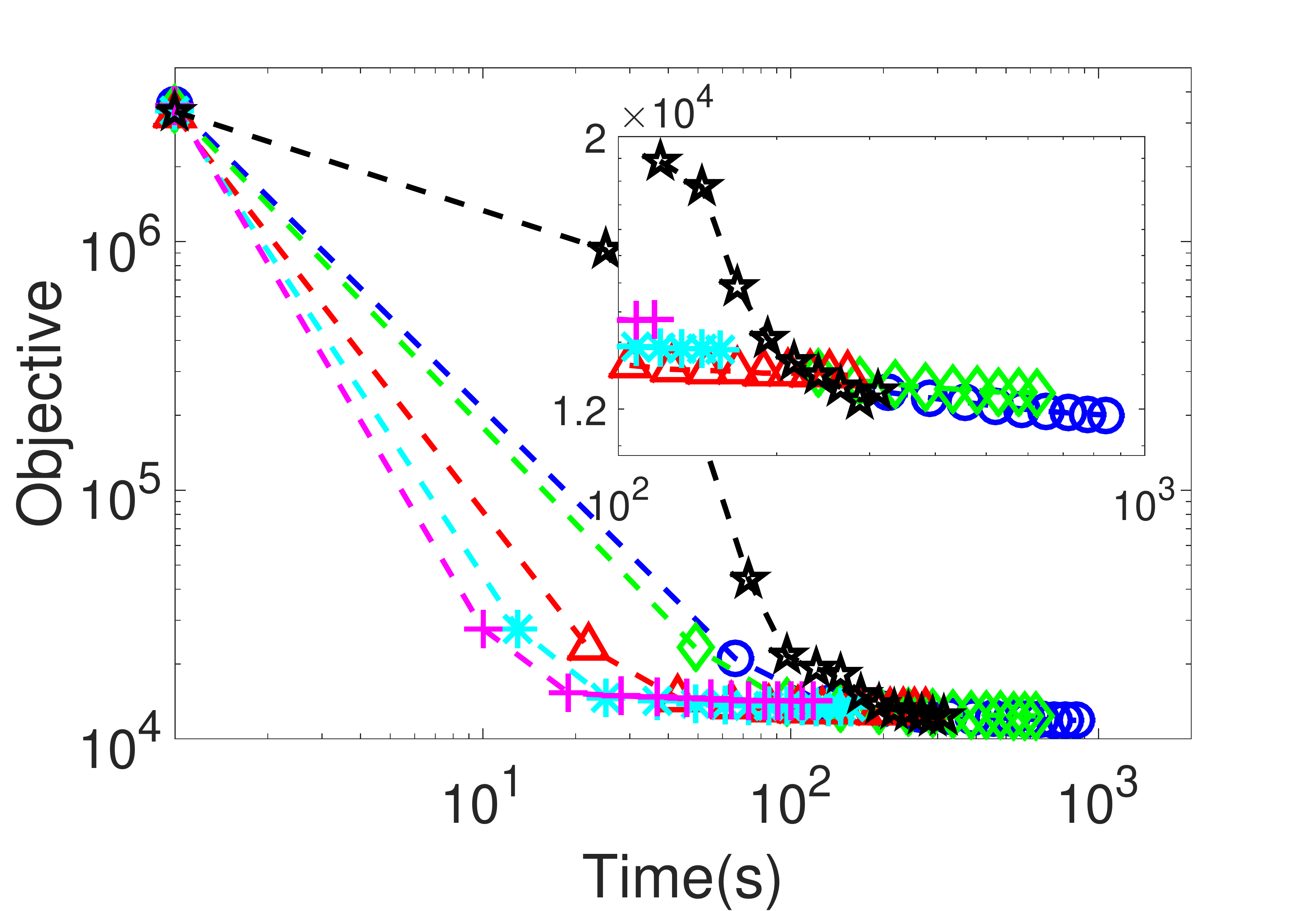}
  \vspace*{1mm}
\end{subfigure}

\caption{Convergence comparison between the-state-of-art method~\cite{heide2015fast} and the proposed method with different subsampling rate, all of which are performed on fruit dataset. Convergence is evaluated by monitoring the objective value of Eq.~\ref{eq:CSCmodel} on training images versus iterations and time, respectively.}
\label{fig:subsampleResult}
\end{figure}

\subsection{Subsampling Strategy}

{\bfseries Convergence.} Comparisons of the convergence between the
proposed method (SBCSC) and the state-of-the-art batch-mode
algorithm~\cite{heide2015fast} are shown in
Fig.~\ref{fig:subsampleResult} (the comparison method uses a similar number of iterations as ours to reach convergence). A different selection of the
subsampling rate reveals that the proposed strategy will slightly
influence the convergence and the training objective of the
minimization problem. Specifically, the more subsampled, the
relatively slower convergence and the higher objective will be
obtained. On the other hand, small subsampling rate will significantly
accelerate the computation process, where $10\%$ subsampling achieves
about $6 \times$ speedup over the not subsampled spatial domain solver
and a $2 \times$ speedup over state-of-the-art Frequency-domain solver
for one iteration. We observe that a subsampling rate between $p=0.1$
and $p=0.2$ delivers empirically good enough convergence in our
settings, as well as achieving at least $3 \times$ speedup. In
general, the proposed method with various subsampling rates converges
at around 10-12 iterations in all testing cases, acting similar to the
competing methods.

In summary, the convergence behaviors of the proposed
algorithm is only slightly influenced by the subsampling strategy
within the testing subsampling rates. Comparing to the
state-of-the-art frequency solver, the proposed stochastic
spatial-domain solver with a subsampling rate of $0.1$ reduces the
computation time by a factor of two for the tested example. Specifically,
SBCSC takes 170 seconds and the comparable method takes 350 seconds for
14 iterations on a Core i7 PC. The
robustness of the proposed algorithm is evaluated by additional
experiments. Please see supplementary materials for reference.


\subsection{Online Learning}

\begin{figure}[h]
\centering
\begin{subfigure}{0.45\textwidth}
  \includegraphics[width=1\linewidth]{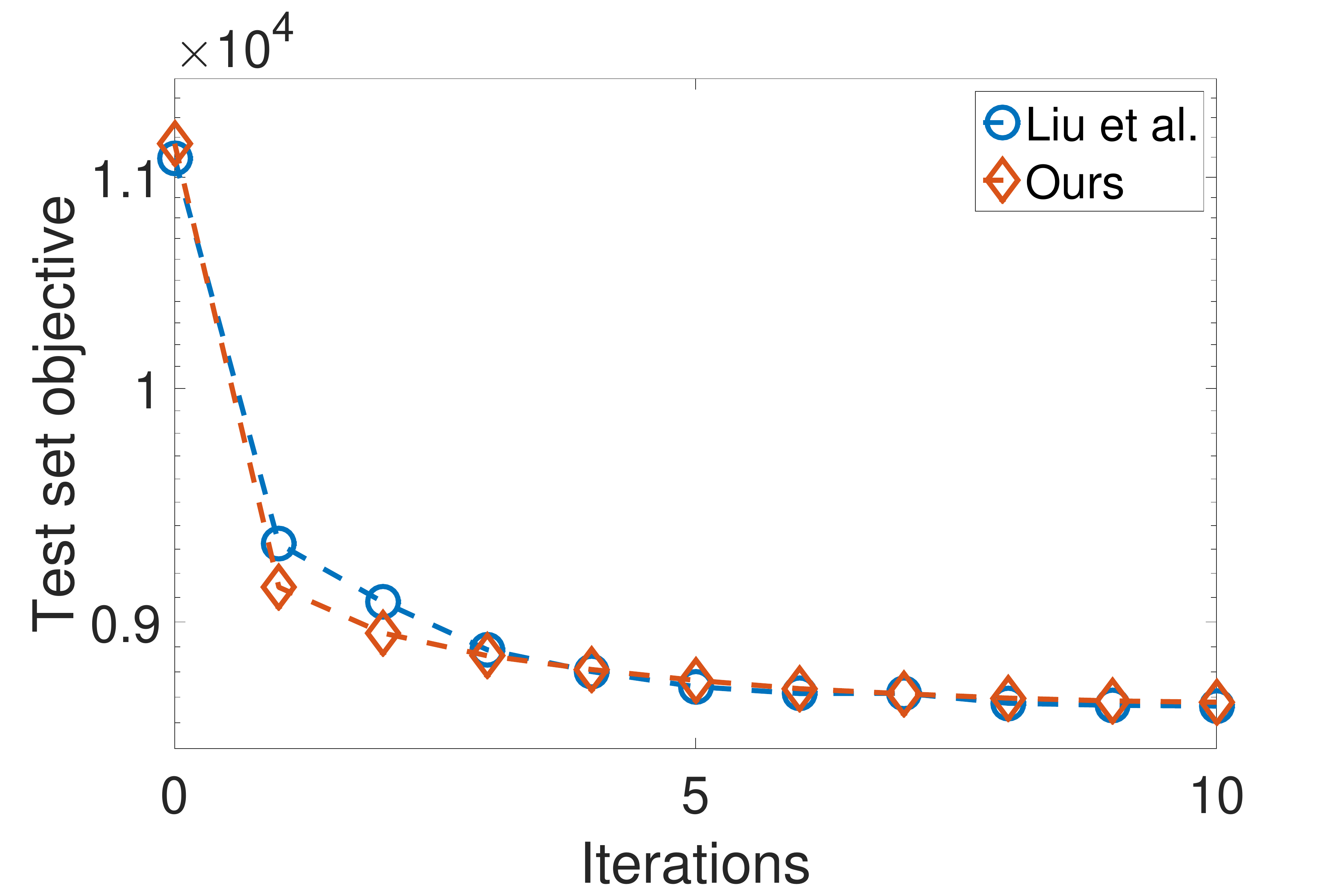}
\end{subfigure}
\begin{subfigure}{0.45\textwidth}
  \includegraphics[width=1\linewidth]{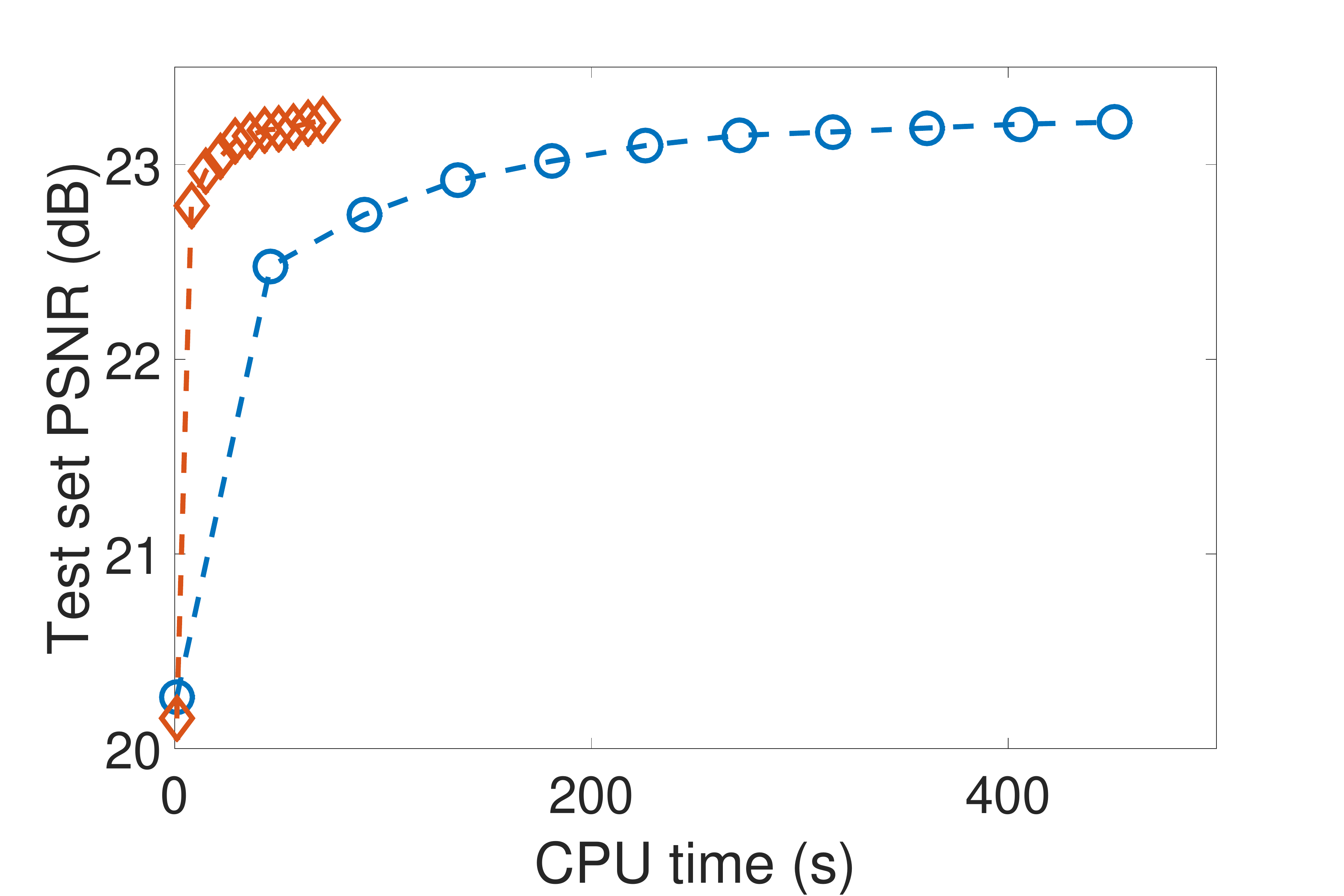}
\end{subfigure}

\caption{The experiments are performed on fruit dataset, and each iteration randomly choose one samples from the training datasets. Top: Convergence of the test set objectives (objective value of Eq.~\ref{eq:CSCmodel} on testing datasets) for our method (SOCSC) and the current online approach~\cite{liu-2018-first}. Bottom: Testing PSNR with respect to execution time. While the quality of the output is comparable, our method achieves $6 \times$ speedup.}
\label{fig:onlineSmall}
\end{figure}

{\bfseries Convergence.} Unlike the batch-based learning approaches
which evaluate its convergence by monitoring the objective value on
training datasets, a common way to evaluate the learning process of
online learning model is to monitor its objective value on test
datasets. In Fig.~\ref{fig:onlineSmall} we plot the objective values
against the iteration number for the proposed method (SOCSC) and a
recent online frequency-domain CSC method~\cite{liu-2018-first} (both
 approaches use Matlab built-in functions only). In
the same figure, we also keep track of the capability of the updated
filters during the learning process to sparsely represent the test
images, which is demonstrated by the time evolution of PSNR (PSNR is the 
peak signal-to-noise ratio, measuring the difference between the 
reconstructed signal and the original one). These two
approaches stop at optimum positions with similar objective values.
The final PSNR values for both methods also reveal a similar
reconstruction performance of the learned filters. In terms of runtime
comparison, however, the proposed method runs at least $6$ times faster
than the comparison method. Specifically, SOCSC takes 70 seconds and the comparable method takes 440 seconds on a Core i7 PC to process all 10 training images. The supplement shows additional comparisons for the other datasets.


\begin{figure}[h]
\centering
\begin{subfigure}{0.45\textwidth}
  \includegraphics[width=1\linewidth]{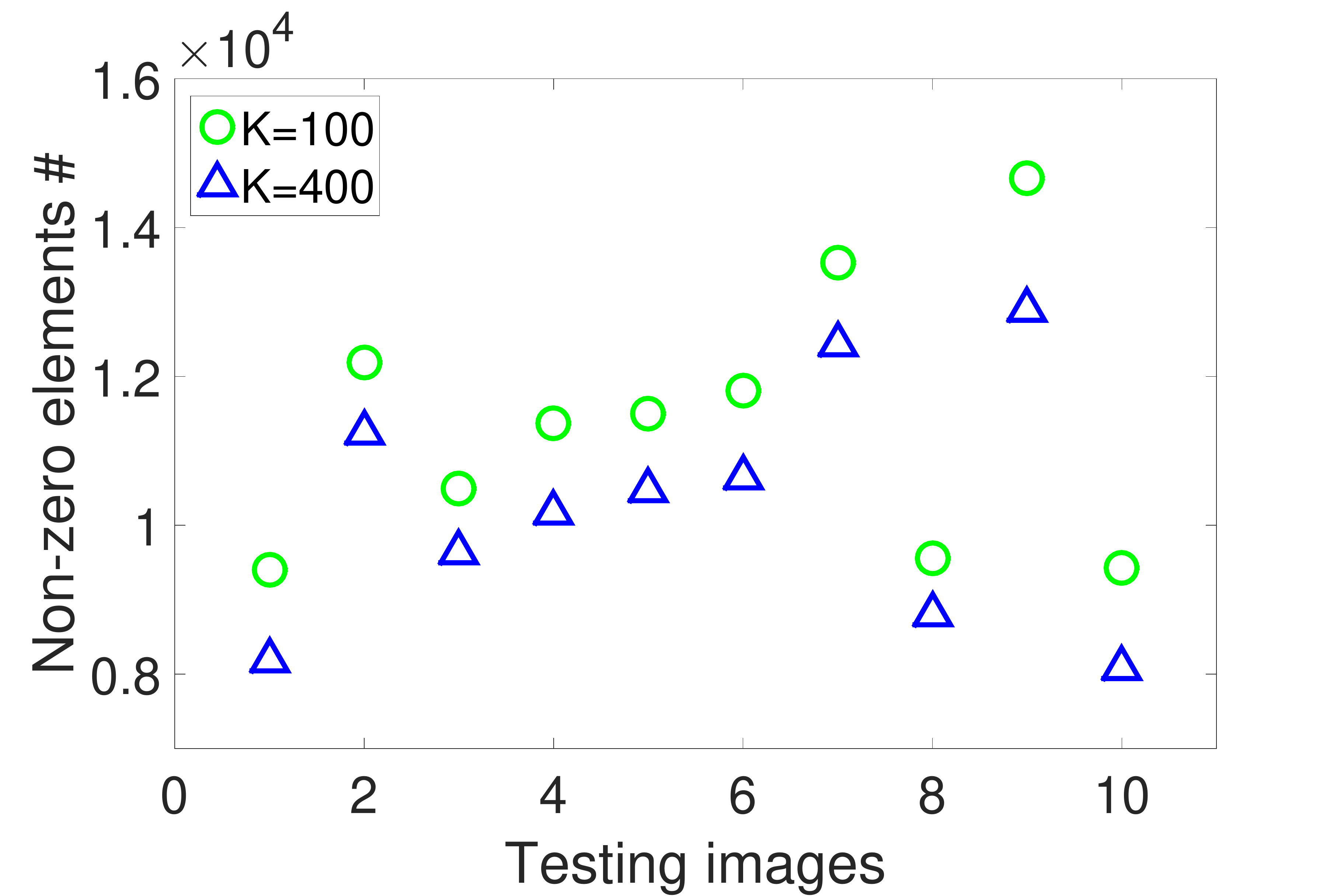}
\end{subfigure}
\begin{subfigure}{0.45\textwidth}
  \includegraphics[width=1\linewidth]{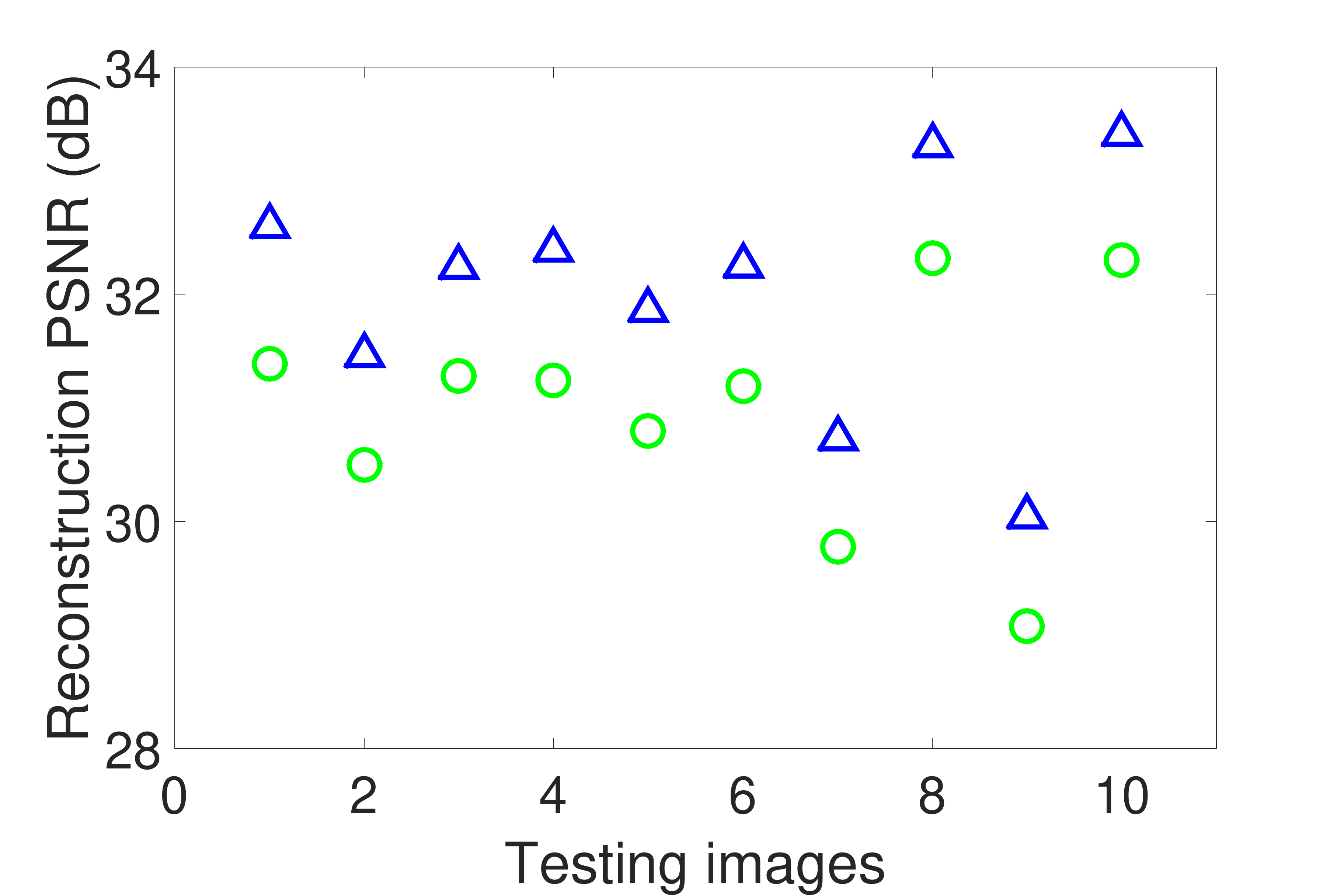}
\end{subfigure}

\caption{Top: number of non-zero elements in the codes for different images (on average $0.2\%$ of the variables are non-zero when $K=100$). Bottom: PSNR between the reconstructed images and the original ones with under-complete and over-complete dictionaries, respectively.}
\label{fig:overVSunder}
\end{figure}

\begin{figure*}[h]
\begin{minipage}{0.4\textwidth}
\begin{subfigure}{1\textwidth}
    \centering
  \includegraphics[width=0.75\linewidth]{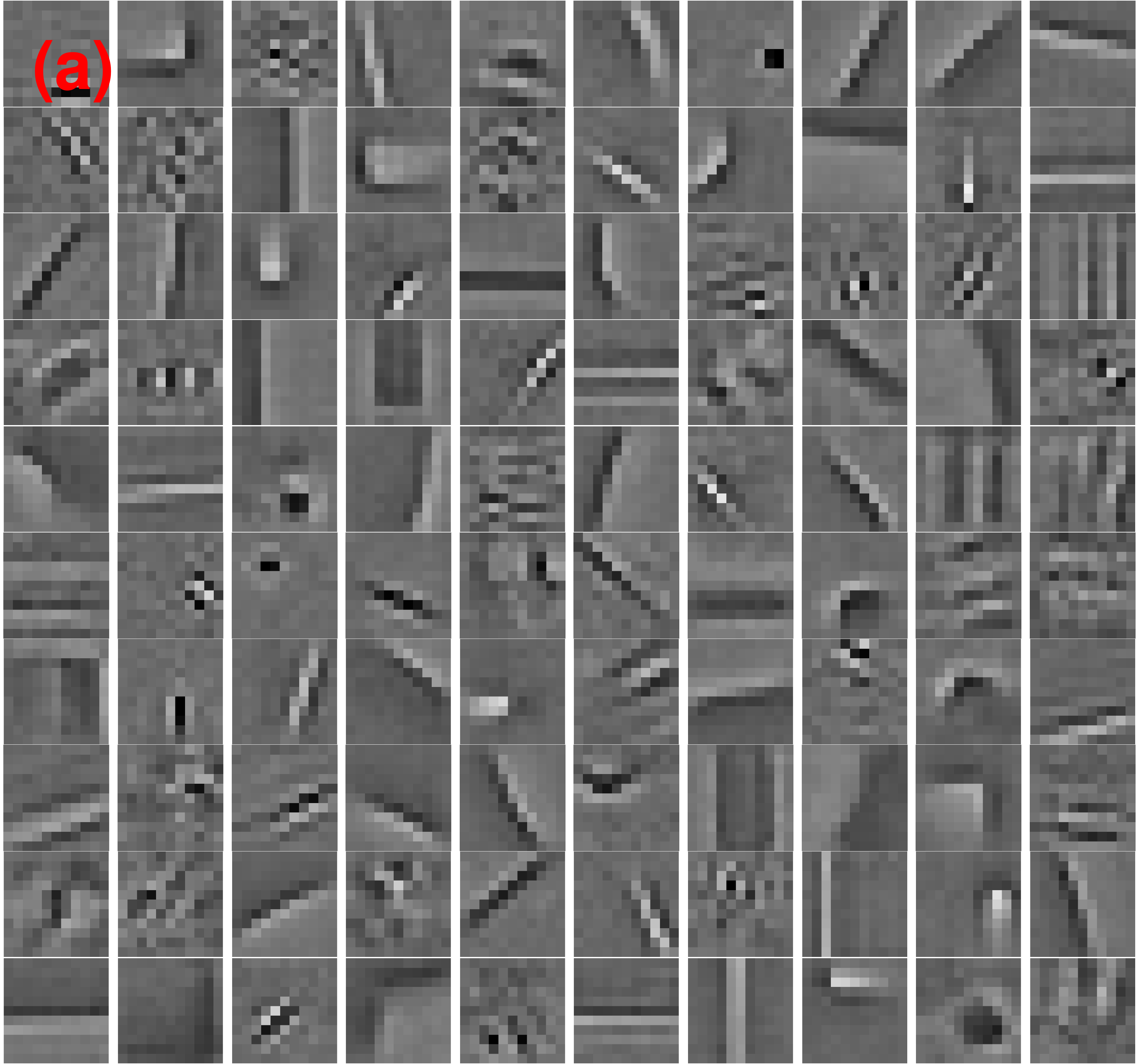}
  \vspace*{2mm}
\end{subfigure}

\begin{subfigure}{1\textwidth}
    \centering
  \includegraphics[width=0.75\linewidth]{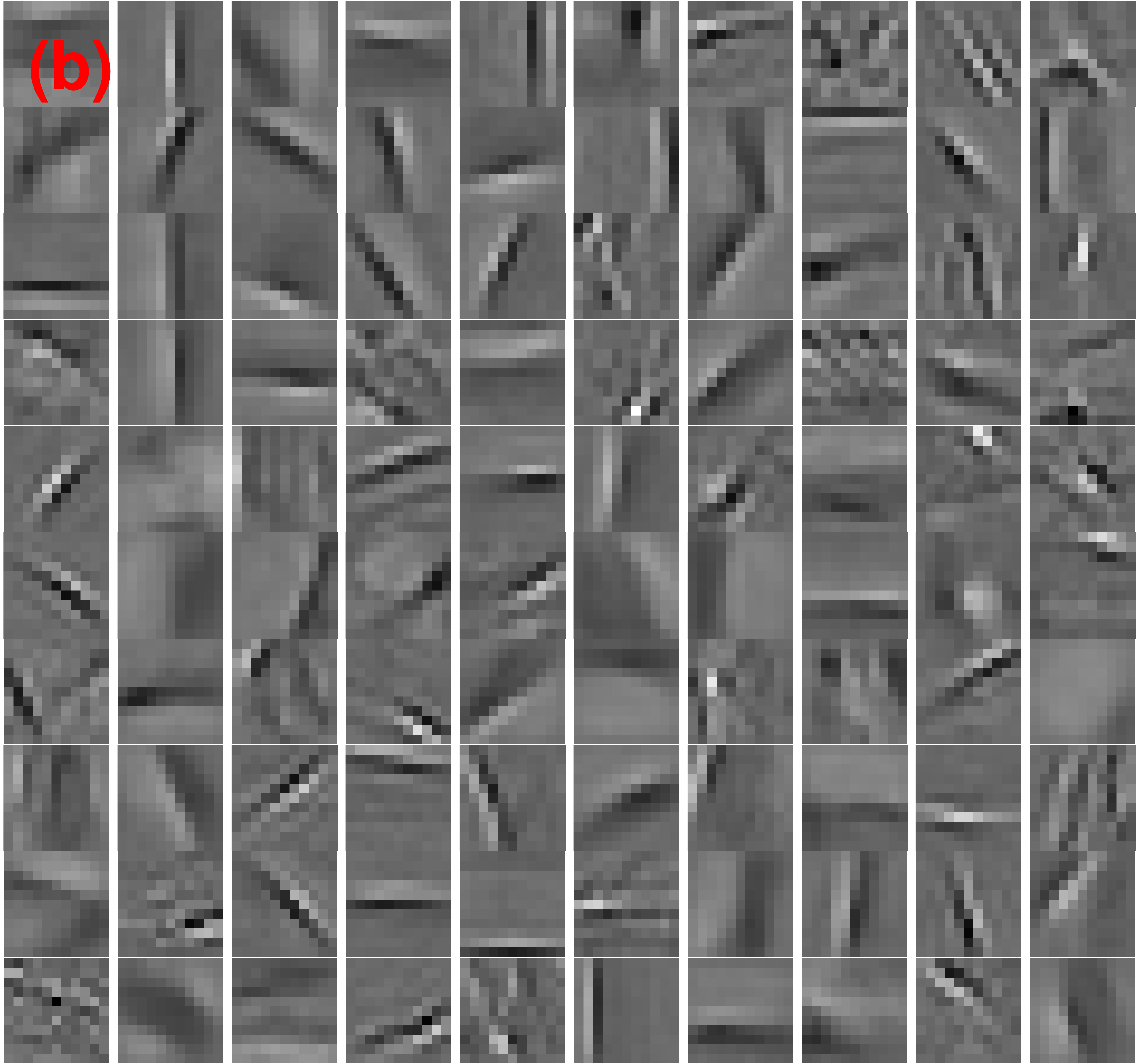}
\end{subfigure}
\end{minipage}
\begin{minipage}{0.6\textwidth}
\centering
\includegraphics[width=0.97\linewidth]{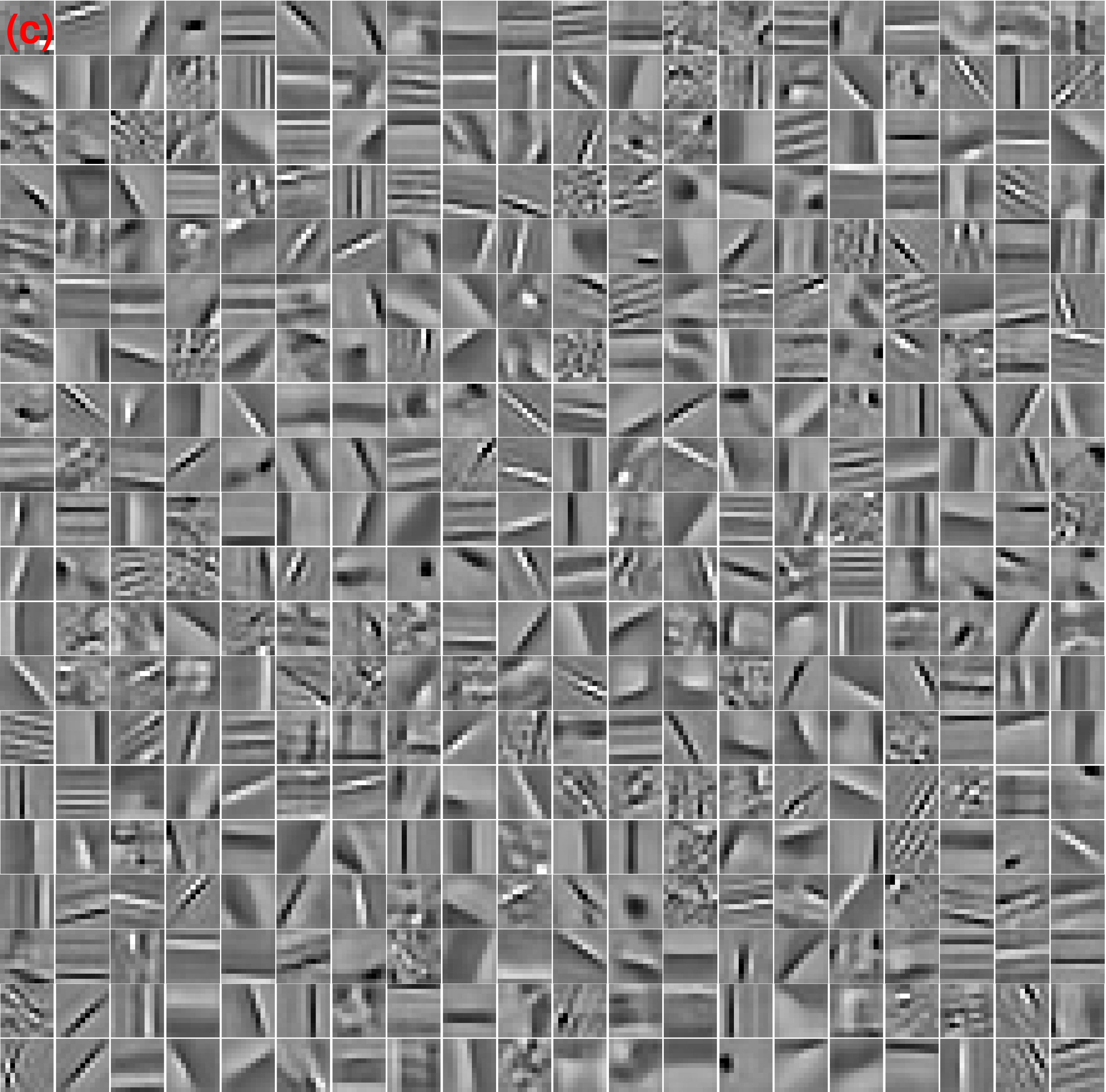}
\vspace*{2mm}
\end{minipage}
\begin{minipage}{1\textwidth}
\centering
\includegraphics[width=0.9\textwidth]{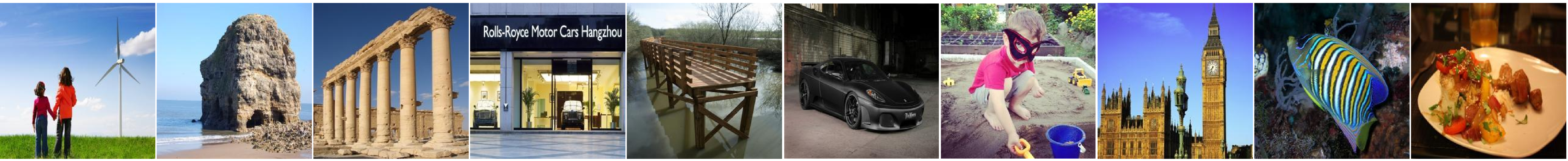}
\vspace*{2mm}
\end{minipage}
\begin{minipage}{1\textwidth}
    \centering
    \resizebox{0.8\linewidth}{!}{
        \begin{tabular}{|c||c|c|c|c|c|c|c|c|c|c|}
            \cline{1-11}
            Image & 1 & 2 & 3 & 4 & 5 & 6 & 7 & 8 & 9 & 10 \\
            \hline
            PSNR by (a) & 29.58 & 28.19 & 29.44 & 29.63 & 28.89 & 29.33 & 28.13  & 30.14 & 27.42 & 30.89 \\
            \hline
            PSNR by (b) & 29.63 & 28.22 & 29.57 & 29.90 & 29.12 & 29.59 & 28.05  & 30.17 & 27.53 & 31.08 \\
            \hline
            PSNR by (c) & \textbf{30.24} & \textbf{28.34} & \textbf{29.95}  & \textbf{30.30} & \textbf{29.43} & \textbf{29.96} & \textbf{28.24} & \textbf{30.57} & \textbf{27.72} & \textbf{31.67} \\
            \hline
        \end{tabular} }
\end{minipage}
\caption{Top: Filters learned from large-scale datasets by our method (SOCSC) and the comparable online method~\cite{liu-2018-first}. Bottom: $10 ~ 256 \times 256$ testing images and their corresponding reconstruction quality in the image inpainting application. (a) The under-complete dictionary ($11 \times 11 \times 100$) learned by~\cite{liu-2018-first}; (b) The under-complete dictionary ($11 \times 11 \times 100$) learned by SOCSC. (c) The over-complete dictionary ($11 \times 11 \times 400$) learned by SOCSC. These under-complete dictionaries, mainly composed of Gabor-like filters, can be seen as a subset of the represented over-complete dictionary, which contains a number of extra low contrast image features.}
\label{fig:overCompleteDic}
\end{figure*}

{\bfseries Over-complete dictionary.}
Learning over-complete dictionary (number of the dictionary is more
than its degrees of freedom) from small datasets would cause
overfitting issues, which may contain quite a few data-specific
filters, and therefore limit the ability to generalize the filters to
other data (we verify this explanation in the supplement). The
proposed online-based learning strategy (SOCSC) can overcome this
issue by scaling the model up to arbitrary sample sizes.

We demonstrate this ability on 1000 image patches with the size of
$100 \times 100$, and learn an $11 \times 11 \times 400$ over-complete
dictionary, which is shown in Fig.~\ref{fig:overCompleteDic}. For a
visual comparison, we also show $100$ learned filters generated by the
same algorithm and another 100 filters generated
by~\cite{liu-2018-first}. As can be observed, both of the approaches
learn visually similar under-complete dictionary, while the proposed
method takes $6 \times$ less training time than the comparison
method. The learned over-complete dictionary is composed of
the Gabor-like image features as represented in the under-complete
dictionaries, as well as a number of low contrast features which are
not typical for under-complete dictionaries. This additional
feature information would play an essential role to reveal an improved sparse representation of
the natural images. The numerical comparisons of number of non-zero elements and its corresponding reconstruction PSNR for testing images ($10 ~ 256 \times 256$ images presented in Fig.~\ref{fig:overCompleteDic}) are shown in Fig.~\ref{fig:overVSunder}. Here, we define the non-zero elements as the codes whose coefficient is no less than $0.1$. We could observe that at all times, using over-complete dictionary leads to a sparser representation of the images, roughly $8\%-10\%$ reduction on the non-zero elements. Meanwhile, it achieves dramatically improved reconstruction quality, over $1$ dB on average.

We further demonstrate the effectiveness of the over-complete
dictionary in the application of image inpainting,
which refers to reconstructing a full image from partial
measurements. A numerical comparison of the reconstruction quality is
shown in the bottom of Fig.~\ref{fig:overCompleteDic}. The reconstructions
are performed on $50\%$ randomly observed images, with $\lambda = 0.4$ and
$50$ ADMM iterations for all cases. Obtained PSNR values are averaged on
5 trials. The over-complete filters learned by the proposed
method achieves significantly improved reconstruction quality on partially 
observed images in terms of the PSNR value.

\begin{figure}[h]
\centering
\begin{subfigure}{0.45\textwidth}
  \includegraphics[width=1\linewidth]{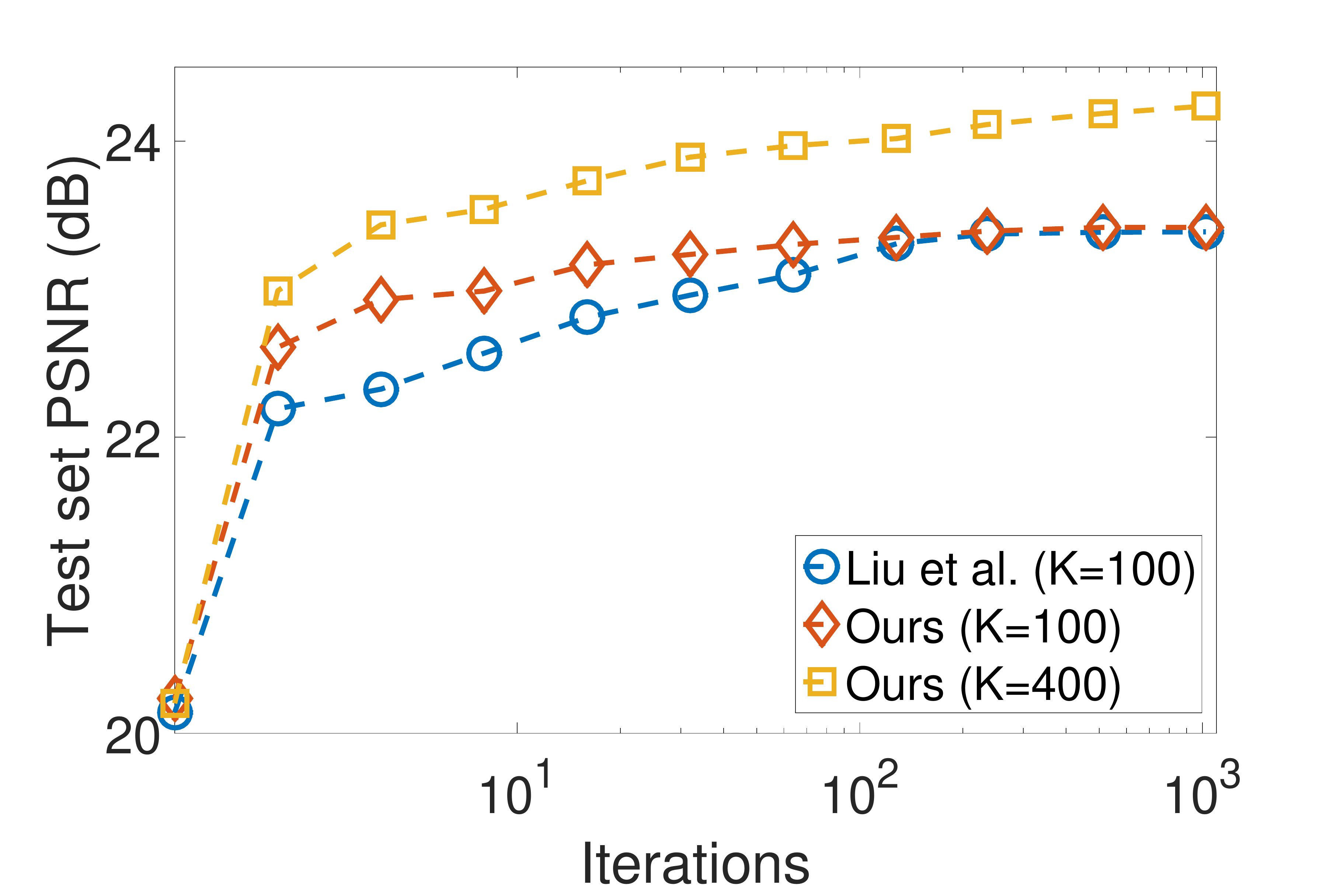}
\end{subfigure}
\begin{subfigure}{0.45\textwidth}
  \includegraphics[width=1\linewidth]{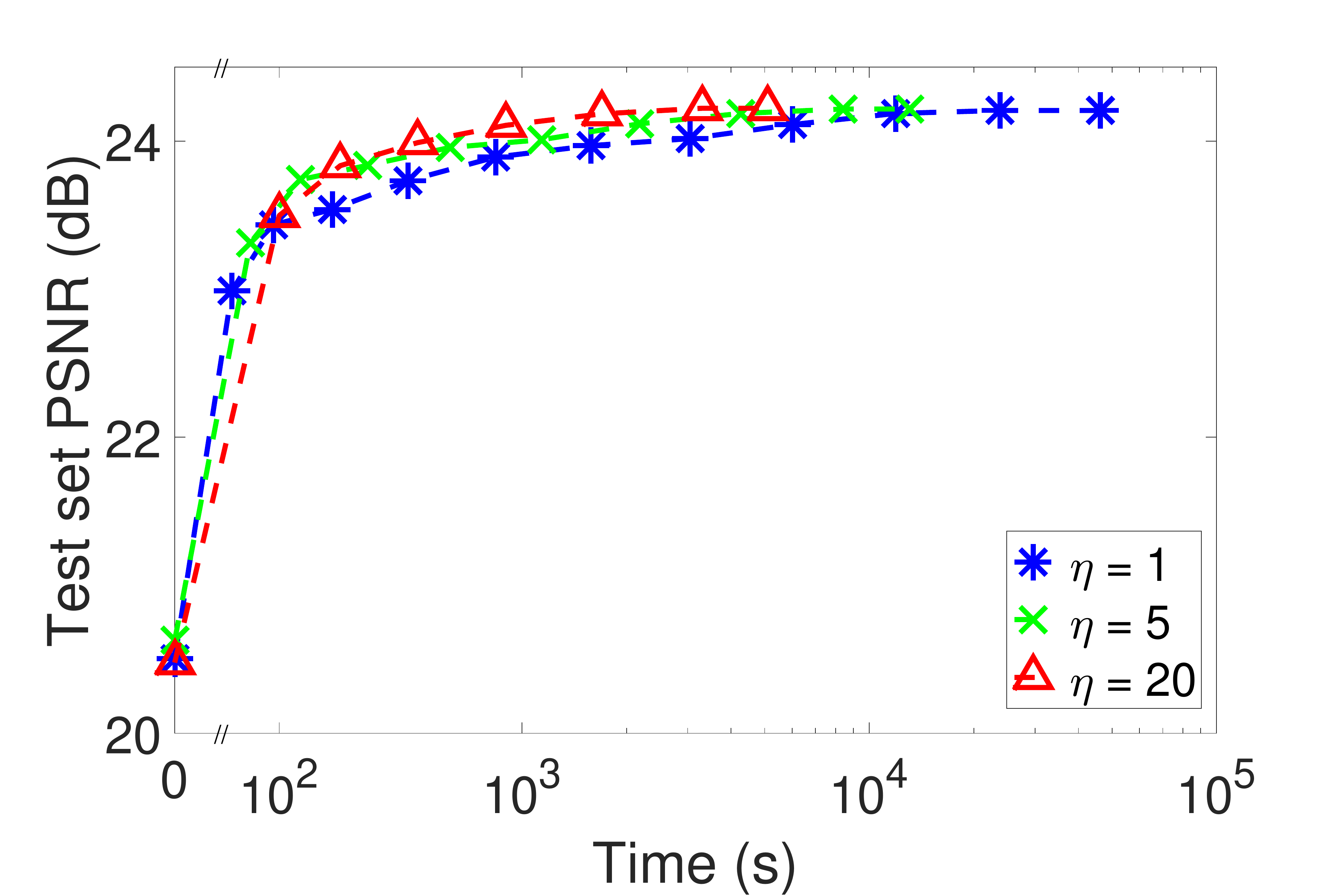}
\end{subfigure}

\caption{Top: Testing PSNR for the comparable method~\cite{liu-2018-first} with $K=100$, and our method (SOCSC) with $K=100$ and $K=400$, respectively. Every iteration draws a single image from those 1000 image patches. Bottom: Testing PSNR for SOCSC ($K=400$) with varying values of $\eta$. The learned filters are examined on the test sets every $2^i$ iterations and also at the last iteration, where $i=0,1,\dots$. Note that all the results are generated by a single-core program.}
\label{fig:overComDicAndMinibatch}
\end{figure}

We also observe a bottleneck revealed by the under-complete dictionary
in the online-based CSC model. The top of Fig.~\ref{fig:overComDicAndMinibatch} demonstrates that no more
apparent progress could be observed when the number of training images
is higher than a specific value for both of the online approaches
($K=100$). However, owing to more abundant filters, learning
over-complete dictionary overcomes this bottleneck, and it shows a
considerable improvement in the PSNR of image representations. All 
presented experimental results imply that the number of filters and 
number of training samples are both essential in the CSC model.

{\bfseries Mini-batching.} In practice, a mini-batching strategy would
be preferred in order to gain advantages from modern parallel
computing architectures. This is also a standard extension to
stochastic optimization algorithms~\cite{Takac2013, PCDM, SCSG}. We
denote the mini-batch size as $\eta$. In the proposed online algorithm
(SOCSC), the time complexity for one step dictionary update will not
increase linearly with the increase of $\eta$. Concretely speaking,
updating $\code$ can be implemented by caching the Cholesky
decomposition, and one computation of the matrix factorization can be
applied to all of the currently selected batches. Herein, the
complexity for doing updating $\code$ $\eta$ times all together is
cheaper than $\eta$ times the complexity of updating one $\code$. In
addition, the time complexity for updating $\filter$ is not linearly
affected by the value of $\eta$, which will be executed only once in
each training step regardless of $\eta$. One exception is the update
of surrogate matrices which has a complexity that is linear in
$\eta$. However, this step is not dominant in the runtime.

The runtime comparisons for various mini-batch sizes are shown in the
bottom of Fig.~\ref{fig:overComDicAndMinibatch}. Note that larger
$\eta$ will result in a smaller number of iterations to process all
$1000$ samples. The plots show that a larger mini-batch size will
generally lead to a greater progress in first few training steps,
though it takes additional running time and memory. Overall,
mini-batched update provides a more runtime efficient learning
process in the online-based CSC model, and according to the obtained
experimental results, $\eta=20$ achieves one order of magnitude
speedup over $\eta=1$ to reach a a comparable level of
convergence. Since computing sparse codes is a data-independent
process, this mini-batched approach can be further accelerated in a
distributed-computing system.


\section{Conclusions}
In this work, we present a novel stochastic subsampling strategy for
solving the CSC problem in the spatial domain. This method exploits 
the sparsity of the over-parameterized model and improves the runtime performance over the prior
frequency-domain solvers, which applies to both batch mode and
online-learning mode.  The proposed algorithm, for the first time,
demonstrates the feasibility that tackling the CSC problem in spatial
domain while still holding, or even improving the runtime
efficiency. Since the subproblem of updating the code is a highly
sparse LASSO, other specific optimization strategies can be applied to
further accelerate the computation, for instance the idea proposed
in~\cite{johnson2017stingycd}, which solves the LASSO problem by
coordinate descent and skips unnecessary updates using the method of
safe screening~\cite{ghaoui2012Swfe}. It is worth emphasizing that
Frequency-domain methods cannot benefit from these kinds of speedup
strategies.


We have also shown the capability of the developed online algorithm to
learn representative and meaningful over-complete dictionary from
arbitrary large datasets, and the availability of the dictionary is
further verified by the application of image inpainting. It can be
foreseen that this capability has widespread applications in audio
and image related tasks, and higher dimensional signal processing.


\section{Acknowledgements}
This work was supported by King Abdullah University of Science and Technology as part of VCC center baseline funding.

\bibliographystyle{eg-alpha-doi}
\bibliography{egbib}   

\end{document}